\documentclass[superscriptaddress, aps, prl, twocolumn]{revtex4-2}

\usepackage{dsfont}
\usepackage{upgreek}
\usepackage{amsfonts}
\usepackage{amsmath}
\usepackage{graphicx}
\usepackage{dcolumn}
\usepackage{bm}
\usepackage[colorlinks]{hyperref}
\hypersetup{allcolors= {blue}}
\usepackage{braket}
\newcommand{\orcid}[1]{\href{https://orcid.org/#1}{\includegraphics[width=7pt]{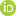}}}

\begin{document}

\title{Time-symmetry and topology of the Aharonov-Bohm effect}

\author{Yakir Aharonov}
\affiliation{Schmid College of Science and Technology, Chapman University, Orange, California 92866, USA}
\affiliation{Institute for Quantum Studies, Chapman University, Orange, California 92866, USA}
\affiliation{School of Physics and Astronomy, Tel Aviv University, Tel Aviv 6997801, Israel}

\author{Ismael L. Paiva}
\affiliation{H. H. Wills Physics Laboratory, University of Bristol, Tyndall Avenue, Bristol BS8 1TL, United Kingdom}
\affiliation{Faculty of Engineering and the Institute of Nanotechnology and Advanced Materials, Bar-Ilan University, Ramat Gan 5290002, Israel}

\author{Zohar Schwartzman-Nowik}
\affiliation{School of Computer Science and Engineering, The Hebrew University, Jerusalem 91904, Israel}
\affiliation{Faculty of Engineering and the Institute of Nanotechnology and Advanced Materials, Bar-Ilan University, Ramat Gan 5290002, Israel}

\author{Avshalom C. Elitzur}
\affiliation{Institute for Quantum Studies, Chapman University, Orange, California 92866, USA}
\affiliation{Iyar, The Israeli Institute for Advanced Research, POB 651 Zichron Ya'akov 3095303, Israel}

\author{Eliahu Cohen}
\affiliation{Faculty of Engineering and the Institute of Nanotechnology and Advanced Materials, Bar-Ilan University, Ramat Gan 5290002, Israel}

\begin{abstract}
The Aharonov-Bohm (AB) effect has been highly influential in fundamental and applied physics. Its topological nature commonly implies that an electron encircling a magnetic flux source in a field-free region must close the loop in order to generate an observable effect. In this Letter, we study a variant of the AB effect that apparently challenges this concept. The significance of weak values and nonlocal equations of motion is discussed as part of the analysis, shedding light on and connecting all these fundamental concepts.
\end{abstract}

\maketitle

In classical physics, deterministic evolution in time combined with the Newtonian picture of point particles acted upon by forces provides a clear explanation of the dynamics of all physical systems. In quantum mechanics, while more subtleties are present, Schr\"odinger's unitary evolution also gives a mathematically elegant picture of the dynamics of closed quantum systems. However, quantum measurements drastically modify this description, implying a non-unitarity often referred to as the ``collapse of the wave function.'' This, in turn, is associated with the measurement problem~\cite{zurek2003decoherence,schlosshauer2005decoherence} and various counter-intuitive notions, e.g., single particle nonlocality~\cite{tan1991nonlocality,hardy1994nonlocality,greenberger1995nonlocality,bjork2001single,dunningham2007nonlocality,paraoanu2011realism,fuwa2015experimental,georgiev2020analysis}.

It seems, at first, that collapse challenges the time-symmetric, reversible notion of Schr\"odinger's unitary evolution. However, Aharonov, Bergmann, and Lebowitz~\cite{aharonov1964time} showed that this is not the case when one augments the initial boundary condition of a quantum system with a final one. This has led to the development of the \textit{two-state vector formalism} (TSVF)~\cite{aharonov2002two, aharonov2005quantum, aharonov2008two,aharonov2014measurement}, which accords with other modern discussions in which past and future are treated on equal footing~\cite{cramer1986transactional,kastner2013transactional,sutherland2008causally,silberstein2018beyond,price2012does,price2015disentangling,wharton2018new,oreshkov2015operational,drummond2020retrocausal,di2021arrow,hardy2021time}. The TSVF ascribes the quantum measurement's outcome to two wavefunctions: one (preselected) proceeding forward in time from the source to the detector and the other (postselected) going in a backward direction. This approach offers a unique perspective for many quantum phenomena, such as single~\cite{georgiev2020analysis,aharonov2013quantum,aharonov2021dynamical,elitzur2018nonlocal} and multipartite~\cite{resch2004extracting,lundeen2009experimental,aharonov2015weak,aharonov2016quantum,aharonov2018completely} quantum nonlocality as quantum effects stemming from the time-symmetrization of the quantum formalism. Furthermore, the TSVF provides an original approach for addressing the measurement problem and the emergence of classicality~\cite{aharonov2002macroscopic,cohen2017quantum,pan2019weak}. More interestingly, under special combinations of pre- and postselected states, novel phenomena emerge that \textit{a posteriori} are obliged by the standard formalism as well~\cite{aharonov1988result, aharonov2013quantum}---yet, they had not been predicted without the TSVF.

However, there is a conceptual trap: How can one verify a pre-/retrodiction about a quantum state that prevails \textit{between} measurements? Weak measurement \cite{aharonov1988result} has been devised to bypass this pitfall, and it turned out to be helpful in various scenarios~\cite{dixon2009ultrasensitive, turner11, susa2012optimal, dressel2013strengthening, jordan2014technical, pang2014entanglement, alves2015weak, harris2017weak, pfender19, cujia19, fang2021weak, huang2021amplification, paiva2022geometric}. Yet, its statistical nature in some of the implementations, and the toll it typically takes in the form of a tiny but non-zero disturbance, have led to debates regarding its ontological role as revealing a fundamental trait of the measured quantum system \cite{vaidman2017weakvalue}. In addition, the ``weak reality''~\cite{vaidman2017weak, aharonov2018weak}, prevailing during the intermediate period between initial and final states, often outlines a perplexing picture regarding the past of the system~\cite{vaidman2013past,danan2013asking,vaidman2014tracing,englert2017past,peleg2019comment,englert2019reply,paneru2017past}.

The present Letter will examine this picture within a curious scenario involving the Aharonov-Bohm (AB) effect. The past of the electron undergoing the AB effect will be seen to create some tension with the topological nature of the effect. Nevertheless, nonlocal equations of motion within the Heisenberg picture will be shown to provide some explanation.

The rest of the Letter is organized as follows. First, we outline the basics of our particle-based approach and exemplify it using a double-well setup. Next, we describe the AB effect within a double Mach-Zehnder interferometer (MZI) probed by weak measurements. We analyze the validation technique and its implications. The main contribution of this work is, then, a thought experiment where a particle seems to exhibit the AB effect even though its (forward-evolving) wavefunction seems to take only one side of the solenoid and does not encircle it.

\textit{The nonlocal, particle-based approach}---In the proposed approach, based on the Heisenberg picture, the notion of superposition is reinterpreted in the language of operators~\cite{aharonov1970deterministic}, especially modular operators obeying nonlocal equations of motion~\cite{aharonov1969modular}. Consequently, the explanation of interference phenomena can be done in an ontologically distinct way~\cite{aharonov2017finally}. Modular variables consist of observables whose expectation value is a quantity associated with the relative phase between wavepackets. The Heisenberg equations of these operators are nonlocal, meaning that their time evolution is affected by potentials in remote locations. For instance, if the Hamiltonian of a system is written as $H=P^2/2m+V(X)$, the Heisenberg evolution of the operator $e^{iP\ell}$ (we use a system of units in which $\hbar=1$), associated with the modular momentum, is
\begin{equation}
    \frac{d}{dt} e^{iP\ell} = -i[e^{iP\ell},H] = -i [V(X+\ell I)-V(X)] e^{iP\ell}.
    \label{eq:mod-momentum}
\end{equation}
This is the case because $e^{iP\ell} |x\rangle = |x+\ell\rangle$ and, then, $e^{iP\ell} V(X)=V(X+\ell I) e^{iP\ell}$. Observe that, even though $e^{iP\ell}$ is non-Hermitian, quantum observables can be built from it (e.g., using its real part).

This type of nonlocality is called \textit{dynamical nonlocality}~\cite{aharonov1969modular}, in opposition to the most common \textit{kinematic nonlocality}, like the one considered in Bell scenarios. It enables the assumption of a picture where the particle has a definite position (even if it is unknown) while consistently explaining interference phenomena. For a concrete example, modular momentum allows the explanation of interference in the double-slit experiment in this picture with localized particles~\cite{aharonov2017finally}. In fact, if the separation between the slits has a length $\ell$, one can assume the particle passes through one of the slits and, nevertheless, has its dynamics affected by potentials placed in the other. Tightly related to this notion of dynamical nonlocality is the AB effect, where a charge's interference pattern is affected by electromagnetic fields confined to a region it did not traverse.

Now, suppose the preparation of a system (preselection) does not specify the location of a particle. In that case, the information of where it was can be, in some cases, obtained with a second measurement later on, i.e., a postselection. However, as already implied above, postselection scenarios allow a picture containing retrocausality in the time interval between pre- and postselection. In light of this discussion, it seems that dynamical nonlocality may be viewed as a forward-in-time manifestation of retrocausal phenomena within a two-time picture and vice versa. Next, we intend to discuss the interplay between these two ideas. We also address the issue of validating novel predictions entailed by it.

\textit{Exploring dynamical nonlocality within a double well}---We start with a simple setup to illustrate some fundamental notions underlying the main result of this Letter, described in the next section. We consider two interconnected cavities, which create a double-well system. If a particle is on the left cavity, we can associate a state, say, $|L\rangle$ to it. Similarly, the state $|R\rangle$ is associated with a particle on the right-hand side cavity. Observe that this can be seen as a way to study interference, in particular, within a MZI. In fact, such interferometers accommodate wavepackets traveling in two separated paths and interfering at the end, such as a simple MZI, or multiple times in the middle of the experiment, like in nested MZIs. The key assumption that allows this mapping is the already mentioned fact that the two cavities are interconnected, which allows interference between wavepackets in different cavities.

\begin{figure}
    \centering
    \includegraphics[width=\columnwidth]{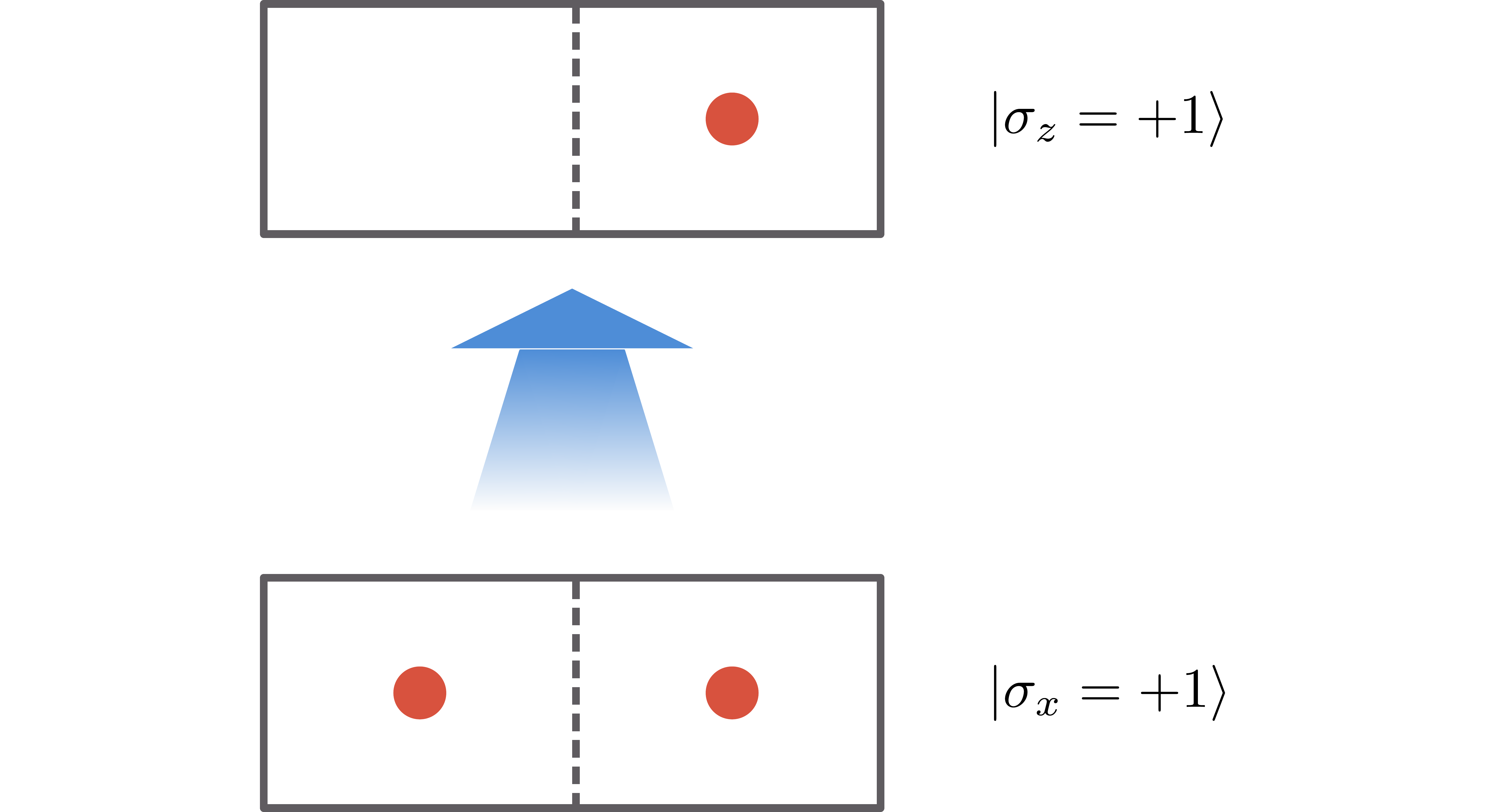}
    \caption{Representation of two interconnected cavities preselected in the state $|\sigma_x=+1\rangle$ and postselected in the state $|\sigma_z=+1\rangle$.}
    \label{fig:boxes}
\end{figure}

It follows that the generic state of a particle, which can be in a superposition of being in both cavities, is $\cos(\theta/2) |L\rangle + e^{i\phi} \sin(\theta/2) |R\rangle$. Since this is just a qubit state, we can use a more familiar notation by identifying $|L\rangle \equiv |\sigma_z = -1\rangle$ and $|R\rangle = |\sigma_z = +1\rangle$, where $|\sigma_z = -1\rangle$ and $|\sigma_z = +1\rangle$ are the eigenstates of the Pauli operator $\sigma_z$ with eigenvalues $-1$ and $+1$, respectively.

Now, suppose that a particle is prepared in the superposition $(|\sigma_z = -1\rangle + |\sigma_z = +1\rangle)/\sqrt{2} = |\sigma_x = +1\rangle$. Later, it is postselected in the final state $|\sigma_z = +1\rangle$, as represented in Fig.~\ref{fig:boxes}. How can one reconcile these fundamentally different boundary conditions? We argue that, within a time-symmetric picture, it might be easier to perceive these two conditions as two properties that the particle possesses: It is well localized in the right cavity, having its Pauli-$z$ determined, and \textit{at the same time} it possesses a well-defined Pauli-$x$ value. To understand in which sense the definiteness of both $\sigma_x$ and $\sigma_z$ is possible, we observe that, in a scenario where each of the eigenstates of $\sigma_z$ corresponds to the presence of the system in a remote location, $\sigma_x$ plays the role of a modular variable~\cite{paiva2022coherence}. In fact, because of the cyclic structure of the problem, the analog of the modular momentum $e^{iP\ell}$ previously mentioned in this Letter is an operator that (up to a global phase) maps $|\sigma_z = -1\rangle$ into $|\sigma_z = +1\rangle$ and vice versa. It can be checked that this can be done with the operator $e^{i\sigma_x \pi/2}$. However, this operator is equivalent to a multiple of the standard $\sigma_x$. To evidence this fact, suppose that, at some intermediate time $t'$, an impulsive potential $V=(1-\sigma_z) V_0 \delta(t-t')/2$, where $V_0$ is a real constant, is applied to the system. The presence of the factor $1-\sigma_z$ means that the potential is applied only on the left-hand side cavity. Nevertheless, according to the Heisenberg equations of motion, the value of $\sigma_x$ \textit{will} change according to
\begin{equation}
    \frac{d}{dt} \sigma_x = -i[\sigma_z V_0  \delta(t-t')] \sigma_x.
\end{equation}
This shows that the dynamics of the modular variable $\sigma_x$ is affected by potentials placed at either location, each associated with $|\sigma_z = -1\rangle$ and $|\sigma_z = +1\rangle$.

Observe that, by symmetry, the same reasoning presented here holds if the pre- and postselected states were inverted, i.e., if the system was preselected in $|\sigma_z = +1\rangle$ and postselected in $|\sigma_x = +1\rangle$, a scenario where the influence of the potential in the dynamics seems even more challenging to comprehend from the perspective of the Schr\"odinger picture. Moreover, it is noteworthy that the modular variable $\sigma_x$ becomes local when the paths recombine, i.e. a modular operator with nonlocal properties at a given instance of time may later correspond to a local property. Next, we discuss a related yet more dramatic case where a solenoid exerts such a nonlocal effect on an electron.

\textit{Dynamical nonlocality within the Aharonov-Bohm effect}---The AB effect~\cite{ehrenberg1949refractive, Aharonov1959} reveals a unique aspect of quantum dynamics that challenges the basic principles of electromagnetism. A current passing through a solenoid, which perpendicularly crosses a MZI, affects the relative phase between an interfering electron's wavepackets. For simplicity, we consider here the case where the current generates a half-integer fluxon inside the solenoid, which flips the sign of the wavepackets' relative phase and, consequently, the electron's exit port on the MZI.

This effect is commonly regarded as a nonlocal interaction between the magnetic flux passing within the solenoid and the electron encircling it~\cite{aharonov2016nonlocality}. However, questions have been raised about the topological and nonlocal nature of the effect~\cite{vaidman2012role, marletto2020aharonov}, including the relationship between this type of nonlocality and kinematic nonlocality, like in Einstein-Podolsky-Rosen-Bell (EPR-Bell) scenarios~\cite{einstein1935can,bohm1957discussion,bell1964einstein,brunner2014bell}. The latter can be studied with general probabilistic methods, which provide bounds for quantities associated with them, allowing for a mathematically rigorous verification in different configurations. In contrast, the AB and related effects present dynamical nonlocality~\cite{aharonov1996non}, which emerges from the dynamical equations of motion. The nature of the latter leads to difficulties in attaining precise operational bounds for its verification: since these bounds are associated with the dynamical equations, it seems that a device-independent approach cannot easily indicate their presence. That being said, in the standard semiclassical treatment of the AB effect, it can be shown that the only quantities that change in a gauge-invariant way while the electron (in a superposition of wavepackets) encircles the solenoid are quantities associated with the electron's modular velocity. In fact, the dynamical equations imply that the modular velocity of the electron has an abrupt, discontinuous change when the line connecting the ``center of mass'' of the packets crosses the solenoid~\cite{aharonov2004effect,kaufherr2014test,paiva2022coherence}.

Now, the line separating kinematics and dynamics is blurry because it depends on the basic framework used to study the theory. The interplay between kinematic and dynamical nonlocalities, therefore, warrants deeper investigation. 

\begin{figure}
    \centering
    \includegraphics[width=\columnwidth]{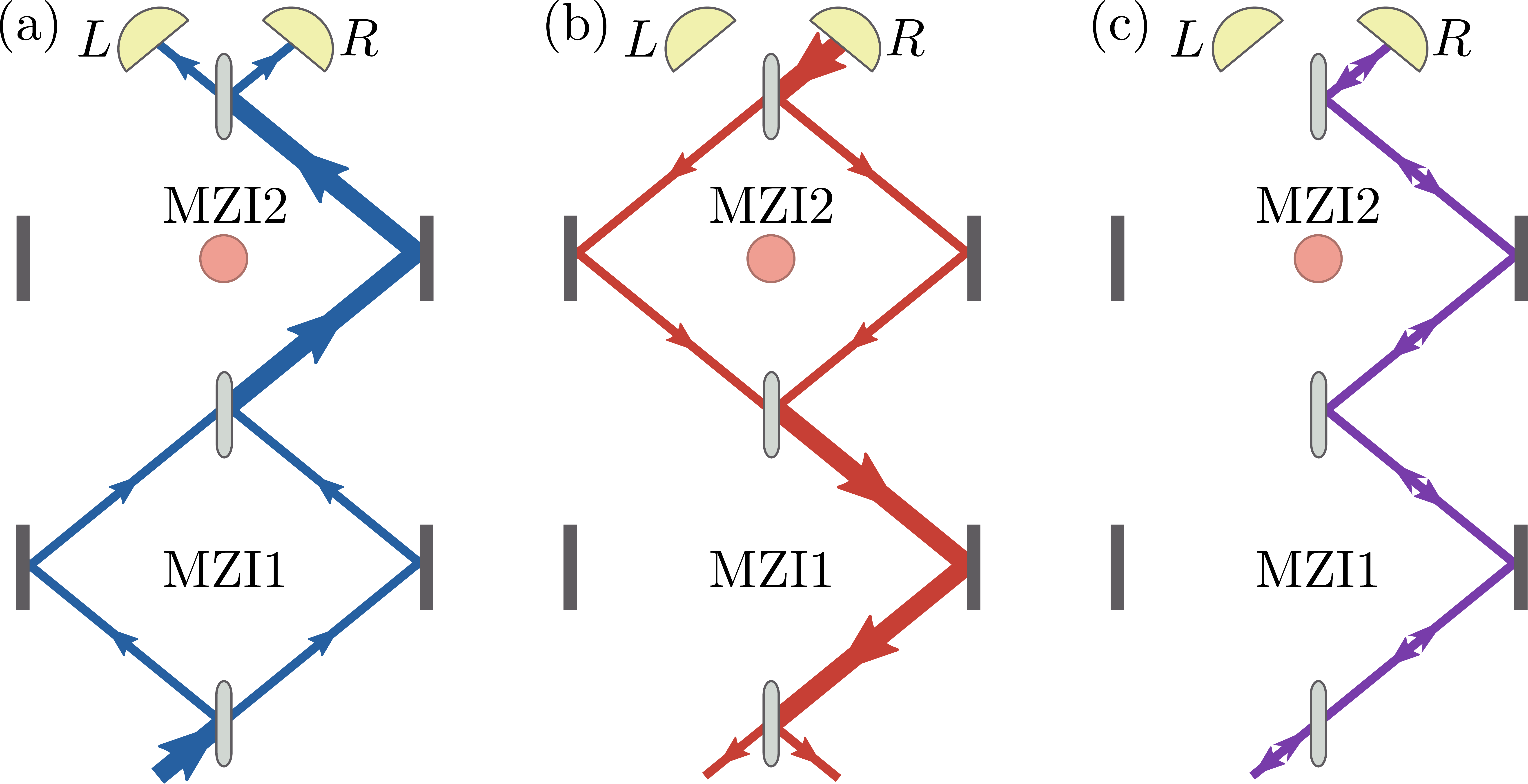}
    \caption{Pre- and postselected states traversing the proposed setup. (a) Forward time evolution. (b) Backward time evolution. (c) Overlap between forward and backward time evolution as seen from the perspective of the two-state vector formalism.}
    \label{fig:dmzi}
\end{figure}

Consider the double MZI shown in Fig.~\ref{fig:dmzi}. While MZI1 is standard, MZI2 encircles a solenoid with a half-integer fluxon. The trajectory of an electron traveling this double interferometer is represented in Fig.~\ref{fig:dmzi}(a). Because it has a definite trajectory in MZI2, one might expect that the presence of the solenoid would exert no effect on the electron's dynamics.

Although this is the case when one considers the standard evolution of the system, the time-symmetric picture presents a peculiar addition. Within the weak reality, the particle's position is determined by both pre- and postselection. Hence the second wavefunction, evolving backward in time, does encircle the solenoid. The system thus travels through regions where the weak value of the position does not vanish. The backward evolution (assuming a postselection of port $R$) is illustrated in Fig.~\ref{fig:dmzi}(b). As a result, between the pre- and postselection, one may assign to the electron a trajectory illustrated in Fig.~\ref{fig:dmzi}(c). From this \textit{combined} perspective, the electron seems to only travel on the right-hand side of MZI2 and, yet, be affected by the solenoid, as can be deduced from its traversal of the right rather than the left arm of MZI1. This assertion can be experimentally proven by performing weak position measurements of the electron while traversing the two MZIs given the prescribed pre- and postselected states.

However, one may ask how this result can be understood in a standard (forward-in-time) picture. For that, modular variables come in handy. Similarly to the above discussion about the double well, besides the position (associated with $\sigma_z$), it is possible to attribute a relevant modular variable (associated with $\sigma_x$) to the electron. Thus, while the solenoid does not affect the electron's position, it does modify its relevant modular variable, which subsequently influences the interference later on.

\textit{Weak measurement disturbance}---One subtle issue will now be addressed. The weak measurements used in MZI1 to collect information about the electron's position spoil the interference pattern that obliges it to take only the right path in MZI2. For this reason, a small portion of all electrons traverses MZI2's ``forbidden'' left arm, thereby creating the observed bias in the final measurements. Indeed, the amount of information gained by these weak measurements is proportional to the number of photons traversing MZI2's left arm~\cite{aharonov2014foundations}. Can one then explain away the correlation between the postselection and the weak pointer shifts to this disturbance?

Theoretically, we can make the coupling strength negligibly small and the ensemble size increasingly large such that the fraction of electrons traversing the left arm of the second MZI goes to zero. Nevertheless, a strong correlation between the final strong measurements and the initial weak measurements will ensue. We shall now analyze this point in more detail.

Assume, as above, that the wavefunctions of an electron traversing the left and right arms are represented by $|\sigma_z=-1\rangle$ and $|\sigma_z=+1\rangle$, respectively. The weak measurement in MZI1 is, therefore, a $\sigma_z$ measurement and can be achieved using the following interaction Hamiltonian, which couples the measured electron and the measuring pointer during some time interval $[0,\tau]$:
\begin{equation}
    H_{\text{int}} = g(t) \sigma_z P,
\end{equation}
where $P$ is the momentum of the pointer and $\int_0^\tau g(t)dt=g_0 \ll \Delta$, with the latter being the standard deviation of the pointer's wavefunction $\Phi(q)= e^{-q^2/4\Delta^2}$.

Under the influence of this Hamiltonian, the initial state of the combined system, proportional to $|\sigma_x=+1\rangle e^{-q^2/4\Delta^2}$, becomes
\begin{equation}
    \begin{aligned}
        e^{i g_0 \sigma_z \otimes P} & |\sigma_x=+1\rangle \otimes \frac{1}{(2\pi \Delta^2)^{1/4}} \int dq \; e^{-q^2/4\Delta^2} |q\rangle \approx \\
        &A \left(|\sigma_x=+1\rangle \otimes \int dq \; e^{-q^2/4\Delta^2} |q\rangle \right. \\
        & \ \ \ \left. - \frac{g_0}{2\Delta^2} |\sigma_x=-1\rangle \otimes \int dq \; q e^{-q^2/4\Delta^2} |q\rangle \right),
    \end{aligned}
\end{equation}
where $A$ is a normalization constant. After reading an outcome $q_0$ off the measuring device's pointer, the state of the system of interest becomes
\begin{equation}
    B \left(|\sigma_x=+1\rangle - \frac{g_0 q_0}{2\Delta^2} |\sigma_x=-1\rangle \right),
    \label{eq:dist-state}
\end{equation}
where $B$ is a normalization constant. The last equation represents the back-action of the measurement and the emerging probability amplitude to traverse the left arm of MZI2. Since typically $q_0 \lesssim \Delta$, the probability of a flip is usually smaller than $g_0^2/\Delta^2 \ll 1$ (in agreement with a previous analysis conducted by Vaidman regarding the weak trace left by a quantum particle~\cite[Eq. (8)]{vaidman2013past}). Hence, one can choose the size of the pre- and postselected ensemble to be $N \gtrsim \Delta^2$ so that the probability that no flip occurs asymptotically scales as $(1-g_0^2/N)^N \rightarrow e^{-g_0^2}$, which can be arbitrarily close to zero. Nevertheless, as indicated in Ref.~\cite{vaidman2013past}, repeating the weak measurements $N \gtrsim \Delta^2$ within the same pre- and postselected scenario leads to a total shift of a pointer which is comparable to the full presence of the electron in one of the arms of the first MZI, based on the accumulation of weak values there. Alternatively, this can be understood as a consequence of interpreting weak values of projection operators as probabilities~\cite{hosoya2010strange, vaidman2017weakvalue}, being in this case 0 and 1 in the forbidden and traversed arms, respectively. As a result, in the limit of large $N$ and correspondingly small coupling strength, the number of electrons traversing the left arm can be made negligibly small, and yet the weak value, as weakly measured in MZI1, implies the existence of the AB effect due to the magnetic flux in MZI2. This apparently paradoxical scenario can be understood either as a two-state effect assisted by the postselected wavefunction in the Schr\"odinger picture or via a dynamically nonlocal equation of motion within the Heisenberg picture. In a subsequent paper~\cite{aharonov:inprep:untitled}, we present an equivalent experiment involving only projective measurements to prove this point.

\textit{Discussion}---The current exploration was motivated by a thought experiment where the Aharonov-Bohm effect seems to have an empirical consequence, even though the electron's (forward-in-time) wavefunction does not encircle a magnetic flux. This has led to a more general observation of the foundations of quantum mechanics.

As shown by the TSVF, it is possible to incorporate a future boundary condition (postselected state) in a consistent and redundancy-free manner to describe systems in quantum theories. In this two-time picture, the state of a system at a given moment is affected by both pre- and postselected states. While this view allows for simple derivations of systems' properties at intermediate times, it may lack a clear description of how systems dynamically evolve and even some of the needed intuition since the pre- and postselected states are fundamentally different. To address these issues, we have shown that it is also possible to understand quantum mechanics as a theory that exhibits both kinematic (Bell-like) and dynamic (AB-like) nonlocal phenomena, which radically differs from classical theory. These phenomena do not lead to causality violations thanks to quantum uncertainty~\cite{cohen2020praise}. In particular, dynamical nonlocality seems to manifest the system's dependence on future configurations in a two-time picture.

The emerging picture is quite simple and self-consistent: The electron can be seen as a particle traversing a definite path rather than a wave superposed over two trajectories. Yet, it has nonlocal properties allowing it to undergo the proposed variant of the AB effect.
We believe that further implications of these ideas should be investigated in a search for a better understanding of the dynamical aspects of quantum systems.

\begin{acknowledgments}
Y.A. thanks the Federico and Elvia Faggin Foundation for support. I.L.P acknowledges support from the ERC Advanced Grant FLQuant. E.C. was supported by the Israeli Innovation Authority under Projects No. 70002 and No. 73795, by the Pazy Foundation, by the Israeli Ministry of Science and Technology, and by the Quantum Science and Technology Program of the Israeli Council of Higher Education. This research was supported by the Fetzer-Franklin Fund of the John E. Fetzer Memorial Trust and by Grant No. FQXi-RFP-CPW-2006 from the Foundational Questions Institute and Fetzer Franklin Fund, a donor-advised fund of Silicon Valley Community Foundation.
\end{acknowledgments}

\bibliography{bibliography}

\end{document}